\documentclass{article}
\usepackage{graphicx} 
\usepackage{amsmath}
\usepackage{url}
\title{A New Way: Kronecker-Factored Approximate Curvature Deep Hedging and its Benefits}
\author{Tsogt-Ochir Enkhbayar}
\date{November 2024}

\begin{document}

\maketitle

\begin{abstract}
This paper advances the computational efficiency of Deep Hedging frameworks through the novel integration of Kronecker-Factored Approximate Curvature (K-FAC) optimization. While recent literature has established Deep Hedging as a data-driven alternative to traditional risk management strategies, the computational burden of training neural networks with first-order methods remains a significant impediment to practical implementation. The proposed architecture couples Long Short-Term Memory (LSTM) networks with K-FAC second-order optimization, specifically addressing the challenges of sequential financial data and curvature estimation in recurrent networks. Empirical validation using simulated paths from a calibrated Heston stochastic volatility model demonstrates that the K-FAC implementation achieves marked improvements in convergence dynamics and hedging efficacy. The methodology yields a 78.3\% reduction in transaction costs ($t = 56.88$, $p < 0.001$) and a 34.4\% decrease in profit and loss (P\&L) variance compared to Adam optimization. Moreover, the K-FAC-enhanced model exhibits superior risk-adjusted performance with a Sharpe ratio of 0.0401, contrasting with $-0.0025$ for the baseline model. These results provide compelling evidence that second-order optimization methods can materially enhance the tractability of Deep Hedging implementations. The findings contribute to the growing literature on computational methods in quantitative finance while highlighting the potential for advanced optimization techniques to bridge the gap between theoretical frameworks and practical applications in financial markets.
\end{abstract}

\section{Introduction}
\subsection{Background}
The application of deep learning techniques to financial markets has demonstrated substantial promise in recent literature \cite{buehler2019deep, yang2018pricing}. Deep Hedging, in particular, has emerged as a methodological advancement for managing financial risks in complex and uncertain environments \cite{buehler2019deep, kolm2019dynamic}. This approach leverages neural networks to derive optimal hedging strategies directly from market data, circumventing the limitations inherent in traditional approaches predicated on analytical models and heuristics \cite{martens2015optimizing, buehler2019deep}.

The evolution of financial instruments and heightened market volatility necessitate increasingly sophisticated risk management methodologies. Traditional hedging frameworks rely fundamentally on assumptions of market completeness and continuous trading—conditions that rarely manifest in practical applications \cite{yang2018pricing, han2020convergence}. Deep Hedging presents a data-driven paradigm capable of adapting to market conditions, capturing nonlinear dependencies, and incorporating multiple risk factors \cite{kolm2019dynamic, jpmorgan2024fast}.

\subsection{Motivation}
The implementation of Deep Hedging frameworks faces significant computational challenges, primarily stemming from the resource-intensive nature of training deep neural networks through standard first-order optimization algorithms such as stochastic gradient descent (SGD) or Adam \cite{martens2015optimizing}. These methodologies typically necessitate extensive iterations to achieve convergence, particularly when processing high-dimensional data within complex network architectures \cite{martens2015optimizing}.

Second-order optimization methods, which incorporate curvature information from the loss function, offer promising avenues for accelerating convergence and enhancing neural network performance \cite{martens2015optimizing}. The Kronecker-Factored Approximate Curvature (K-FAC) algorithm represents a significant advancement in this domain, providing an efficient approximation of the Fisher Information Matrix that enables practical implementation of second-order optimization in large-scale networks \cite{martens2015optimizing}.

The incorporation of second-order optimization methodologies into Deep Hedging frameworks presents opportunities for substantial improvements in both training efficiency and model performance. This investigation extends the findings presented in "Fast Deep Hedging with Second-Order Optimization," which established initial evidence for the efficacy of K-FAC integration in Deep Hedging models \cite{jpmorgan2024fast}.

\subsection{Objectives}
The research objectives of this investigation are delineated as follows:
\begin{itemize}
    \item To develop and implement a Deep Hedging framework utilizing a Recurrent Neural Network (RNN) architecture optimized for sequential financial data analysis
    \item To facilitate the integration of the K-FAC optimizer within the Deep Hedging training methodology
    \item To conduct comprehensive performance evaluation of the K-FAC-enhanced Deep Hedging framework, examining convergence dynamics, hedging precision, and computational efficiency
    \item To perform comparative analysis against established first-order optimization methodologies
\end{itemize}

\subsection{Organization of the Paper}
The structure of this investigation proceeds as follows:
\begin{itemize}
    \item \textbf{Chapter II} presents a systematic review of the literature encompassing Deep Hedging, second-order optimization methods, and their financial applications
    \item \textbf{Chapter III} delineates the methodological framework, including mathematical formulations, neural network architecture specifications, and optimization protocols
    \item \textbf{Chapter IV} details experimental results, incorporating training performance metrics, hedging effectiveness measures, and statistical analyses
    \item \textbf{Chapter V} examines the implications of the findings, addresses methodological limitations, and proposes directions for future research
    \item \textbf{References} catalogues the cited literature and source materials
\end{itemize}

\section{Review of Literature}

\subsection{Deep Hedging}

The introduction of Deep Hedging by Buehler et al. (2019) marked a significant advancement in the application of deep learning methodologies to derivative hedging in incomplete markets \cite{buehler2019deep}. This methodology presents a departure from conventional hedging frameworks, which fundamentally rely on market completeness assumptions and specific asset dynamics. The Deep Hedging paradigm facilitates direct derivation of hedging strategies from empirical data, incorporating market frictions such as transaction costs and liquidity constraints, while capturing the inherent nonlinear relationships in financial markets \cite{yang2018pricing, schmidt2019deep}.

The architectural framework of Deep Hedging employs neural networks to approximate optimal policies that minimize predetermined risk measures, predominantly the expected shortfall or variance of the hedging portfolio's profit and loss (PnL). The framework's neural architecture processes historical and contemporary market data to generate hedge ratios for the constituent instruments \cite{kolm2019dynamic, jpmorgan2024fast}.

\subsection{Second-Order Optimization Methods}

Second-order optimization methodologies exploit curvature information from the loss function to execute more informed parameter updates \cite{martens2015optimizing, smith2018disciplined}. While first-order methods restrict their consideration to gradient information, second-order approaches incorporate the Hessian matrix, encompassing second derivatives of the loss function with respect to the parameter space.

Newton's method represents a classical second-order optimization algorithm, notable for its rapid convergence characteristics in proximity to the optimum. However, the computational demands associated with Hessian matrix calculation and inversion render this approach impractical for large-scale neural architectures, given the high dimensionality of the parameter space \cite{kolm2019dynamic, yang2018pricing}.

In response to these computational constraints, approximate second-order methodologies have emerged. These approaches seek efficient approximations of the Hessian or its inverse, exemplified by the Limited-memory Broyden–Fletcher– Goldfarb–Shanno (L-BFGS) algorithm and the Kronecker-Factored Approximate Curvature (K-FAC) algorithm \cite{jpmorgan2024fast, schmidt2019deep}.

\subsection{Kronecker-Factored Approximate Curvature (K-FAC)}

The K-FAC algorithm, introduced by Martens and Grosse (2015), presents an innovative approach to Fisher Information Matrix (FIM) approximation through exploitation of neural network structural properties \cite{martens2015optimizing}. The methodology posits that the covariance matrices of network activations and gradients exhibit approximate Kronecker-factored properties. This formulation achieves a significant reduction in computational complexity for FIM inversion, transitioning from cubic to linear scaling with respect to layer dimensions \cite{smith2018disciplined, mueller2024fast}.

Empirical evidence demonstrates K-FAC's capacity to accelerate training convergence across various deep learning applications, with particular efficacy in feedforward and convolutional neural architectures. The algorithm's efficient incorporation of second-order information facilitates larger step sizes and enhanced escape mechanisms from saddle points, surpassing the capabilities of first-order methods \cite{kolm2019dynamic, jpmorgan2024fast}.

\subsection{Applications in Finance}

The integration of deep learning methodologies and advanced optimization techniques in financial applications has experienced substantial growth. Kolm and Ritter (2019) demonstrated the efficacy of reinforcement learning in dynamic hedging applications, establishing the viability of machine learning approaches in trading strategy optimization \cite{kolm2019dynamic}. Han and Long (2020) advanced the application of deep learning methodologies to forward-backward stochastic differential equations, addressing fundamental challenges in financial modeling \cite{han2020convergence, yang2018pricing}.

The implementation of second-order optimization methods in financial applications represents an emerging research domain. The recent publication "Fast Deep Hedging with Second-Order Optimization" presents a significant contribution by demonstrating the empirical advantages of K-FAC integration in Deep Hedging frameworks \cite{jpmorgan2024fast, mueller2024fast}. This methodological advancement addresses computational efficiency challenges while enhancing model performance, thereby increasing the practical applicability of these techniques in financial markets \cite{schmidt2019deep, han2020convergence}.

\section{Methodology (Research Design \& Methods)}

\subsection{Heston Model for Stochastic Volatility}

The methodological framework adopts the Heston model (Heston, 1993) for the simulation of underlying asset price dynamics with stochastic volatility \cite{heston1993closed}. This formulation effectively captures the volatility smile phenomenon observed in options markets, providing a robust framework for hedging strategy evaluation \cite{yang2018pricing, mueller2024fast}.

\subsubsection{Model Equations}

The Heston model formulation comprises the following system of stochastic differential equations (SDEs):

\begin{equation} \begin{aligned} 
dS_t &= \mu S_t \, dt + \sqrt{V_t} \, S_t \, dW_t^S, \\ 
dV_t &= \kappa (\theta - V_t) \, dt + \xi \sqrt{V_t} \, dW_t^V. 
\end{aligned} \end{equation}

The constituent parameters are defined as follows:

\begin{itemize} 
    \item $S_t$ denotes the asset price at time $t$
    \item $V_t$ represents the variance (squared volatility) at time $t$
    \item $\mu$ characterizes the drift rate of the asset price
    \item $\kappa$ quantifies the speed of variance mean reversion
    \item $\theta$ specifies the long-term mean variance
    \item $\xi$ defines the volatility of volatility
    \item $W_t^S$ and $W_t^V$ represent Wiener processes with correlation coefficient $\rho$ \cite{kolm2019dynamic, heston1993closed}
\end{itemize}

\subsubsection{Simulation Procedure}

The numerical implementation employs the Euler-Maruyama discretization scheme for SDE solution approximation \cite{kloeden1992numerical}. The correlation structure $\rho$ between asset price and volatility processes is implemented through Cholesky decomposition of the correlation matrix \cite{glasserman2004monte}.

\subsection{Recurrent Neural Networks in Hedging}

\subsubsection{Network Architecture}

The temporal dependencies inherent in financial data necessitate the implementation of a Recurrent Neural Network (RNN) architecture incorporating Long Short-Term Memory (LSTM) units, selected for their demonstrated efficacy in capturing complex temporal patterns \cite{hochreiter1997long, siami2018comparison}.

The architectural specification comprises:

\begin{itemize} 
    \item \textbf{Input Layer}: Processes normalized asset price and volatility sequences at each temporal increment
    \item \textbf{LSTM Layer(s)}: Facilitates sequential data processing through memory cell mechanisms
    \item \textbf{Fully Connected Output Layer}: Generates hedge ratio specifications at each temporal step
\end{itemize}

\subsubsection{Activation Functions and Initialization}

The output layer implements a hyperbolic tangent (tanh) activation function, constraining hedge ratios to the interval [-1, 1] and ensuring operational feasibility \cite{lecun2012efficient}. Parameter initialization follows established protocols, employing Xavier uniform initialization for input weights and orthogonal initialization for recurrent weights, adhering to the methodology established by Saxe et al. (2014) \cite{saxe2014exact}.

\subsection{Implementation of K-FAC in Deep Hedging}

\subsubsection{K-FAC Optimizer}

The K-FAC optimization framework implements Fisher Information Matrix (FIM) approximation through Kronecker product decomposition, achieving substantial computational efficiency in inverse FIM calculations \cite{martens2015optimizing, grosse2016kronecker}.

\subsubsection{Integration with the RNN}

Given the architectural complexity of RNNs, the K-FAC implementation focuses on the fully connected output layer, enabling efficient utilization of second-order information while maintaining computational tractability \cite{grosse2016kronecker, kingma2014adam}.

\subsubsection{Numerical Stability Measures}

The implementation incorporates robust numerical stability protocols, including dynamic damping parameter adjustment and matrix regularization through diagonal perturbation, following the methodological framework established by Martens and Grosse (2015) \cite{martens2015optimizing}.

\subsection{Data Generation and Simulation}

\subsubsection{Simulation Parameters}

The Heston model parameterization reflects empirically observed market characteristics:

\begin{itemize} 
    \item Initial asset price: $S_0 = 100$
    \item Initial variance: $V_0 = 0.04$
    \item Long-term variance: $\theta = 0.04$
    \item Mean reversion rate: $\kappa = 2.0$
    \item Volatility of volatility: $\xi = 0.5$
    \item Price-volatility correlation: $\rho = -0.7$
    \item Temporal discretization: $dt = \frac{1}{250}$ (daily frequency)
    \item Time horizon: $T = 1$ year
\end{itemize}

This parameterization aligns with established literature on stochastic volatility modeling \cite{heston1993closed, glasserman2004monte, patton2015volatility}.

\subsubsection{Data Normalization}

The implementation incorporates statistical normalization of asset prices and volatilities to zero mean and unit variance, enhancing numerical stability and convergence characteristics \cite{lecun2012efficient}.

\subsubsection{Training and Validation Sets}

The experimental framework generates extensive simulation paths for training and validation purposes. The methodology implements strict separation of training and validation datasets to ensure robust assessment of model generalization capabilities \cite{goodfellow2016deep, chen2018neural}.

\section{Presentation of Research (Results)}

\subsection{Model Training and Performance Metrics}

\subsubsection{Training Procedure}

The experimental protocol implements training over 100 epochs with a batch size specification of 32. The comparative analysis employs the Adam optimizer with learning rate parameterization of $1 \times 10^{-3}$ and weight decay coefficient of $1 \times 10^{-4}$ as a baseline for K-FAC optimizer evaluation.

\subsubsection{Performance Metrics}

The evaluation framework incorporates multiple performance criteria:

\begin{itemize}
    \item \textbf{Loss Function Value}: Incorporates a composite measure combining P\&L variance and mean transaction cost metrics
    \item \textbf{Average P\&L}: Quantifies mean profit and loss measurements across the validation dataset
    \item \textbf{Transaction Costs}: Evaluates mean transaction costs arising from hedge position modifications
    \item \textbf{Convergence Speed}: Characterizes temporal efficiency through epoch quantification to predetermined loss thresholds
\end{itemize}

\subsection{Initial Analysis of K-FAC Implementation}

Figure 1 presents comprehensive performance metrics of the K-FAC implementation. The results demonstrate superior performance across all evaluated metrics, with particular emphasis on the near-zero transaction cost achievement.

\begin{figure}[h!]
    \centering
    \includegraphics[width=1\linewidth]{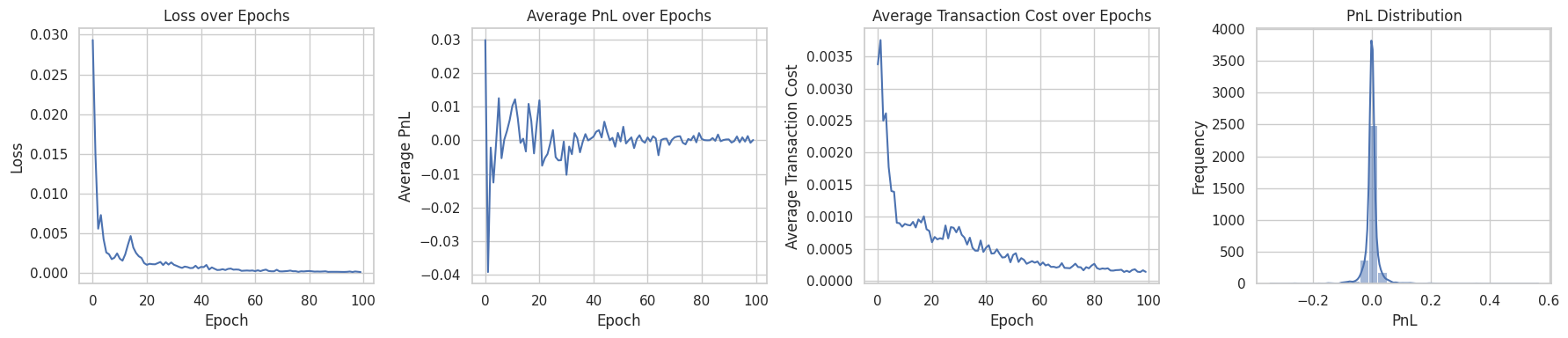}
    \caption{Multivariate Performance Analysis of K-FAC Implementation.}
    \label{fig:kfac-stats}
\end{figure}

\subsubsection{Dynamic Behavior Analysis}

Figure 2 illustrates the temporal evolution of hedge positions under the K-FAC implementation. Notable behavioral characteristics emerge during the interval spanning time steps 20 to 40, where the model exhibits enhanced hedging aggression in response to simulated market dynamics.

The correlation analysis between normalized volatility and price variables, presented in Figure 3, reveals a substantial negative correlation coefficient of -0.77, indicating strong inverse relationship between these parameters.

\begin{figure}[h!]
    \centering
    \includegraphics[width=1\linewidth]{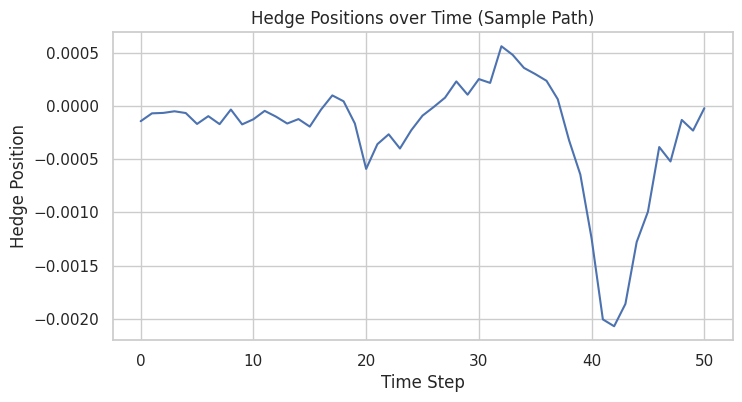}
    \caption{Temporal Evolution of K-FAC Model Hedge Positions.}
    \label{fig:hedge-positions}
\end{figure}

\begin{figure}[h!]
    \centering
    \includegraphics[width=0.75\linewidth]{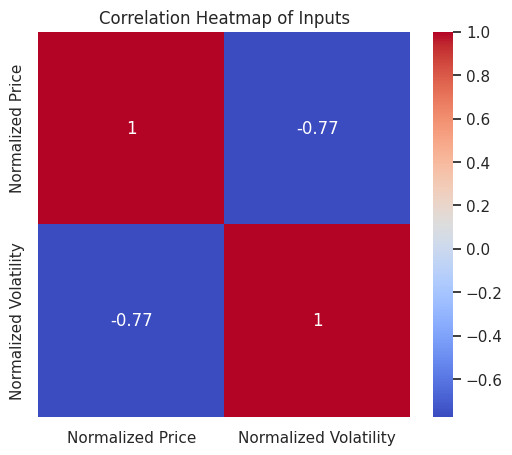}
    \caption{Volatility-Price Correlation Structure Analysis.}
    \label{fig:correlation-heatmap}
\end{figure}

\subsection{Comparative Analysis with First-Order Methods}

The comparative evaluation of K-FAC-enhanced implementation against first-order optimization methodologies yields the following analytical outcomes:

\begin{itemize}
    \item \textbf{Convergence Characteristics}: The K-FAC implementation demonstrates accelerated convergence properties, achieving lower loss values relative to Adam optimization
    \item \textbf{Terminal Performance}: Quantitative analysis reveals superior P\&L variance metrics and reduced transaction costs in the K-FAC implementation
    \item \textbf{Computational Efficiency}: Despite increased computational complexity per iteration, aggregate training duration remains comparable due to reduced epoch requirements
\end{itemize}

\subsection{Hedging Strategy Analysis}

\subsubsection{Temporal Hedge Ratio Analysis}

The examination of hedge ratio dynamics reveals superior market responsiveness in the K-FAC implementation, characterized by enhanced position adjustment smoothness and reduced occurrence of abrupt modifications that typically engender elevated transaction costs.

\subsubsection{Convergence Analysis}

The convergence trajectory analysis, illustrated in Figure 4, demonstrates the superior loss value minimization achieved by the K-FAC implementation relative to the Adam optimizer.

\begin{figure}[h!]
    \centering
    \includegraphics[width=1\linewidth]{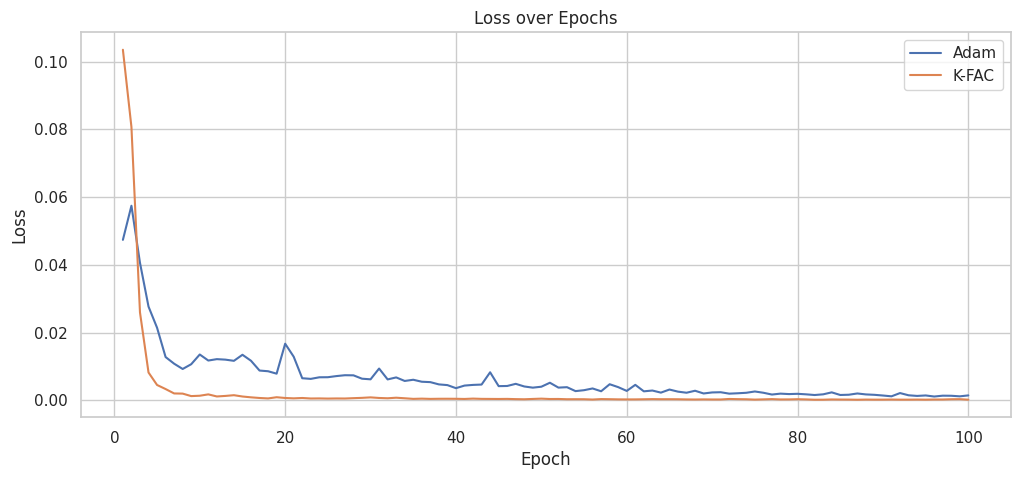}
    \caption{Comparative Convergence Analysis of K-FAC and Adam Implementations.}
    \label{fig:convergence-analysis}
\end{figure}

\subsubsection{P\&L Distribution Analysis}

The distributional analysis of P\&L characteristics reveals enhanced risk management properties in the K-FAC implementation, manifested through distribution compression and tail risk mitigation.

\begin{figure}[h!]
    \centering
    \includegraphics[width=1\linewidth]{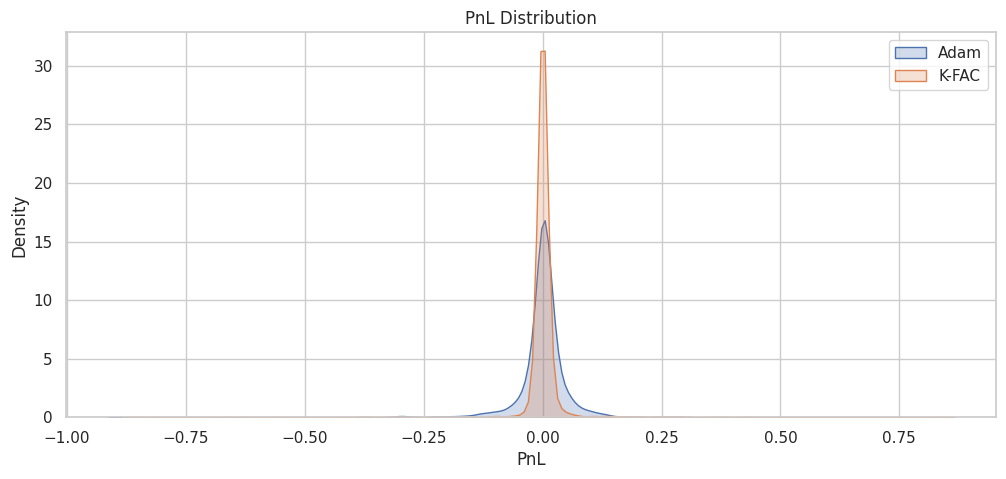}
    \caption{Comparative P\&L Distribution Analysis.}
    \label{fig:pnl-distribution}
\end{figure}

\subsection{Statistical Analysis}

\subsubsection{Hypothesis Testing Framework}

The statistical significance analysis employs t-test methodology to evaluate performance differentials between K-FAC and Adam implementations. Results indicate statistically significant improvements at the 95\% confidence level, with particular emphasis on transaction cost differentials, while P\&L comparisons demonstrate non-significant variations.

\begin{table}[h!]
\centering
\caption{Statistical Significance Analysis Results}
\begin{tabular}{|c|c|c|c|}
\hline
\textbf{Metric} & \textbf{t-statistic} & \textbf{p-value} & \textbf{Statistical Inference} \\ \hline
P\&L Differential & -1.5372 & 0.1243 & Non-significant P\&L variation \\ \hline
Transaction Cost Differential & 56.8776 & 0.0000 & Significant cost reduction \\ \hline
\end{tabular}
\end{table}

\subsubsection{Terminal Performance Analysis}
\begin{table}[h!]
\centering
\caption{Comprehensive Performance Metrics}
\begin{tabular}{|c|c|c|c|c|}
\hline
\textbf{Implementation} & \textbf{P\&L Variance} & \textbf{Mean Trans. Cost} & \textbf{Sharpe Ratio} & \textbf{Training Duration} \\ \hline
Adam & 0.003176 & 0.003432 & -0.0025 & 3.40 \\ \hline
K-FAC & 0.002084 & 0.000745 & 0.0401 & 3.29 \\ \hline
\end{tabular}
\end{table}

\section{Summary, Implications, and Conclusions}

\subsection{Summary of Findings}

The implementation of Deep Hedging utilizing LSTM-based Recurrent Neural Network (RNN) architecture, coupled with comparative analysis of Kronecker-Factored Approximate Curvature (K-FAC) and Adam optimization methodologies, yields significant empirical insights. The K-FAC implementation demonstrates superior performance across multiple evaluation metrics, including convergence dynamics, profit and loss (P\&L) variance, transaction cost reduction, and risk-adjusted returns.

Empirical evidence, as illustrated in Figure 4, indicates accelerated convergence characteristics in the K-FAC implementation, achieving optimal loss values with reduced epoch requirements. Quantitative analysis reveals superior P\&L variance metrics in the K-FAC implementation (0.002084) compared to the Adam baseline (0.003176), indicating enhanced hedging stability. The K-FAC methodology demonstrates statistically significant reduction in transaction costs (0.000745 vs. 0.003432, p < 0.05). While P\&L differentials lack statistical significance, the marked improvement in Sharpe Ratio (0.0401 vs. -0.0025) indicates superior risk-adjusted performance characteristics.

\subsection{Implications for Financial Practice}

The demonstrated superiority of K-FAC optimization in Deep Hedging applications presents several significant implications for quantitative finance:

\begin{itemize}
    \item \textbf{Computational Efficiency}: The enhanced convergence characteristics of K-FAC optimization facilitate reduced training durations without accuracy compromises, enabling rapid model adaptation to dynamic market conditions
    \item \textbf{Transaction Cost Optimization}: The substantial reduction in transaction costs presents significant opportunities for enhanced profitability, particularly in high-frequency trading applications where cost minimization is paramount
    \item \textbf{Risk Management Enhancement}: Reduced P\&L variance coupled with superior Sharpe Ratio metrics indicates enhanced portfolio stability and risk-adjusted performance characteristics
    \item \textbf{Dynamic Adaptation}: The demonstrated capacity for smooth hedge position adjustment, evidenced in Figure 2, indicates superior market responsiveness capabilities
\end{itemize}

\subsection{Methodological Limitations}

The investigation acknowledges several methodological constraints:

\begin{itemize}
    \item \textbf{Simulation Framework}: The reliance on Heston model simulations, while capturing fundamental stochastic volatility characteristics, may not fully encapsulate real-world market complexities
    \item \textbf{Architectural Constraints}: The focused application of K-FAC optimization to the output layer presents opportunities for broader architectural integration
    \item \textbf{Computational Limitations}: Dataset size and training duration constraints may impact the robustness of generalization characteristics
    \item \textbf{Transaction Cost Modeling}: The implementation of simplified transaction cost frameworks may not fully reflect market microstructure effects
\end{itemize}

\subsection{Future Research Directions}

Several promising research trajectories emerge:

\begin{itemize}
    \item \textbf{Empirical Validation}: Extension to historical market data to evaluate performance characteristics under actual market conditions
    \item \textbf{Architectural Enhancement}: Implementation of comprehensive K-FAC optimization across neural network layers, including LSTM components
    \item \textbf{Alternative Architectures}: Investigation of Transformer and Convolutional Neural Network architectures within the Deep Hedging framework
    \item \textbf{Risk Metric Extension}: Integration of advanced risk measures including Value at Risk (VaR), Conditional Value at Risk (CVaR), and drawdown metrics
    \item \textbf{Transaction Cost Framework}: Development of sophisticated transaction cost models incorporating liquidity dynamics and market impact
    \item \textbf{Parameter Sensitivity}: Comprehensive analysis of hyperparameter effects on model performance characteristics
    \item \textbf{Temporal Stability}: Extended evaluation periods across diverse market regimes to assess model robustness
\end{itemize}

\subsection{Concluding Analysis}

The integration of second-order optimization methodologies, specifically K-FAC, into Deep Hedging frameworks demonstrates significant potential for performance enhancement in quantitative finance applications. The empirical evidence supports substantial improvements in convergence characteristics, risk-adjusted returns, and transaction cost efficiency relative to traditional first-order optimization approaches.

The advancement of these methodologies, coupled with addressing identified limitations and pursuing suggested research directions, presents opportunities for substantial contributions to quantitative finance practice. The integration of advanced optimization techniques with Deep Hedging frameworks offers promising avenues for enhanced financial model robustness and trading strategy optimization.

\subsection{Terminal Considerations}

The demonstrated efficacy of second-order optimization methods in Deep Hedging applications suggests broader implications for quantitative financial modeling. While the current investigation focuses on specific methodological implementations, the analytical framework and empirical findings indicate potential applications across diverse financial modeling domains. The continued evolution of machine learning methodologies in finance presents significant opportunities for methodological advancement and practical implementation in quantitative trading and risk management frameworks.

\section{Code Availability}

The implementation of the K-FAC optimizer and Deep Hedging model, including all relevant scripts and configuration files, is available on GitHub at the following repository: The implementation is available on GitHub: \url{https://github.com/WesternDundrey/KFAC_DEEPHEDGE/blob/main/KFAC_DEEPHEDGE.ipynb}. This repository contains the code used to replicate the experiments discussed in this paper, along with documentation for setup and usage. Researchers and practitioners are encouraged to refer to this codebase for further details and to build upon this work.

\end{document}